\documentclass[review]{elsarticle}

\usepackage{lineno,hyperref}
\modulolinenumbers[5]
\usepackage[margin=1in]{geometry}

\usepackage{svg}
\usepackage{enumitem}

\usepackage{array}
\usepackage{rotating}

\usepackage{graphicx}
\usepackage{listings}
\usepackage{makecell}
\usepackage[utf8]{inputenc}
\usepackage[T1]{fontenc}

\usepackage{tikz}
\usetikzlibrary{shapes.geometric, arrows.meta, positioning, fit, backgrounds, calc}

\usepackage{mathpazo} 
\usepackage{caption}
\usepackage[justification=centering]{caption}
\usepackage{setspace}
\usepackage{algorithm}

\usepackage{algpseudocode}

\usepackage[numbers,sort&compress]{natbib}
\biboptions{numbers,sort&compress}

\usepackage{sectsty}
\sectionfont{\fontsize{10}{10}\selectfont}

\journal{Computers \& Security}

\usepackage{amssymb}

\begin{document}

\begin{frontmatter}



\title{Assessing Cyber Risks in Hydropower Systems Through HAZOP and Bow-Tie Analysis}


\author[]{Kwabena Opoku Frempong-Kore}
\ead{kfrem@uis.edu}

\author[]{Rishikesh Sahay}
\ead{rsaha@uis.edu}

\author[]{Md Rasel Al Mamun}
\ead{mmamu@uis.edu}

\author[]{Bell Eapen}
\ead{bpunn@uis.edu}

\address[1]{Department of Management Information Systems, University of Illinois, Springfield, USA}

\begin{abstract}

With the widespread use of software systems in critical infrastructures such as hydropower plants has brought many advantages, yet it has exposed these systems to cyber threats. Cyber risk assessment \& mitigation is important to identify cyber threats and protect these systems from unwanted incidents. 
This paper evaluates and compares the two risk assessment methodologies namely Hazard and Operability Study (HAZOP) and BowTie analysis for identifying cyber induced threats in hydropower systems. 
We selected these two methodologies because they offer a complementary perspective for cyber-safety risk assessment.
Each method is first applied in traditional form to identify hazards, barriers, and threat scenarios arising from accidental causes, then extended to examine how findings change under cyber-induced causation.
The traditional HAZOP identifies 18 deviations across five control parameters; the cyber extension shows how an adversary can coordinate multiple deviations to produce outcomes that conventional safeguards cannot detect.
The BowTie analysis maps preventive and mitigation barriers around a top event; the cyber extension reveals that barriers appearing independently can share network infrastructure a single attacker could compromise, challenging the defense-in-depth assumption.
Together, the two methods provide complementary coverage: HAZOP systematically enumerates what can go wrong, while BowTie shows how barriers provide layered protection.
The cyber extension applied to both exposes assumptions, independent causes in HAZOP and independent barriers in BowTie, that do not hold against a coordinated adversary.
As a result of this study, this paper highlights a practical two-stage approach to adapt established safety methods to identify cybersecurity challenges in hydropower control systems, provides pros and cons of these methodologies, and shows areas of applicability.

\end{abstract}




\end{frontmatter}


\section{Introduction}
\label{sec:introduction}

Hydropower plants are among the oldest and most widely used sources of renewable energy, providing more than 14\% of global electricity generation and close to half of all renewable electricity output worldwide~\cite{IEA2024Hydro}. 
Their operational lifetimes often span decades, during which the underlying control infrastructure has evolved from purely mechanical and electromechanical systems to digitally networked architectures incorporating programmable logic controllers (PLCs), SCADA systems, and remote monitoring capabilities. 
This transition has brought significant operational benefits, including improved efficiency, better integration with grid dispatch systems, and the ability to monitor and control plant functions remotely.
It has also exposed these systems to cyber risks that these systems were not designed to handle originally~\cite{osti_1899145}.

Unlike thermal or nuclear power plants, which have received significant attention from the cybersecurity community, hydropower has historically occupied a quieter corner of the critical infrastructure landscape. 
This is partly because many hydropower facilities are small, geographically dispersed, and increasingly connected to broader grid networks in ways that expand the potential attack surface ~\cite{NREL2024Hydro}. 
It is also due to the physical consequences of a cyber incident at a dam, while potentially severe, that have until recently remained largely theoretical; that is no longer the case.

In April 2025, pro-Russian hackers breached the control systems of the Lake Risevatnet dam in Bremanger, Norway, and remotely opened its floodgates, releasing approximately 7.2 million liters of water over four hours before operators detected the intrusion and regained control~\cite{NorwayDam2025}. 
The attack exploited a web-accessible human-machine interface protected by a weak password~\cite{claroty}.
No injuries or structural damage occurred, largely because the river levels were well below flood capacity at the time. 
The Norwegian authorities classified the incident as an act of hybrid warfare and publicly attributed it to Russia.
The incident was notable not for its technical sophistication but for its simplicity: a single weak credential on an internet-exposed interface was sufficient to achieve physical manipulation of the dam infrastructure.

This is not the first time dam control systems have been targeted.
In 2013, Iranian hackers gained access to the SCADA system of the Bowman Avenue Dam in Rye Brook, New York, through a cellular modem connected to the internet~\cite{DOJ2016Iran}. 
The attacker obtained information about water levels, temperature, and sluice gate status but could not manipulate the gate remotely, as it was manually disconnected for maintenance at the time. 
In 2015 cyber attacks on the Ukrainian power grid demonstrated that adversaries are willing and able to use access to industrial control systems to cause real-world operational disruption in the energy sector ~\cite{CISA2016Ukraine}.

These incidents show that cyber risk to the hydropower infrastructure is a serious issue and it is important to secure these systems.
To strengthen the cybersecurity of hydropower systems, it is important to have an overall view of hydropower systems.
Therefore, the first step is to establish an architecture or framework and identify the System under Consideration (SuC) for cyber risk assessment.
Second step is to perform cyber risk analysis and identify countermeasures.
Moreover, cyber risk analysis should not only consider the risk at the individual component level but also the way it can propagate to other interconnected components and compromise the whole infrastructure.
The components in these hydropower systems are interconnected, therefore the disruption in one part can cause domino effect damaging the whole infrastructure. 
Therefore, having an overall system-of-systems approach is imperative in cyber risk assessment of critical infrastructure, such as hydropower systems.

Still the application of the structured risk assessment approach to this specific domain remains limited. 
Safety-oriented methods such as the Hazard and Operability Study (HAZOP) and BowTie analysis have long histories in the chemical, oil and gas, and nuclear industries, where they are used to systematically identify hazards, map barriers, and inform safety management~\cite{IEC61882}.
However, to apply these methods to cyber physical systems such as hydropower, it requires some adaptation.
Traditional HAZOP assumes that deviations arise from accidental causes; it does not naturally accommodate an intelligent adversary who can coordinate multiple deviations simultaneously.
Traditional BowTie analysis assumes that barriers fail independently; this assumption is difficult to maintain when barriers share a digital infrastructure that a single attacker could compromise.

A small but growing body of work has begun to address this gap. Researchers have applied STPA-Sec, STRIDE, and CORAS to maritime cyber-physical systems, demonstrating how safety-oriented and security-oriented methods can complement each other when applied to the same system~\cite{Sahay2023}. 
Others have explored how established safety analysis techniques, such as FMEA and STPA can be extended to incorporate security considerations \cite{Schmittner2014, Young2014}, and how BowTie analysis can be combined with attack tree methods to address cyber risks in industrial control systems~\cite{Abdo2018}. 
Specifically in the hydropower sector, recent work has focused on developing cybersecurity response and recovery guidance~\cite{PNNL2020Recovery}, developing cybersecurity risk valuation frameworks for hydropower plants ~\cite{sanghvi2023hydropower}, and building situational awareness tools for grid-connected hydropower systems~\cite{NREL2024Hydro}. 
However, systematic application of established safety risk assessment methods, extended to account for cyber threats, to a concrete hydropower control subsystem has not been widely demonstrated.

The main objective of this paper is to address this gap. 
First, we establish a smart hydropower plant as a cyber-physical system within a smart grid environment, define a representative control subsystem (the spillway gate control system), and apply two risk assessment methodologies to that subsystem: HAZOP and BowTie. Each method is first applied in its conventional form to identify hazards, barriers, and threat scenarios arising from accidental causes. 
The analysis is then extended to examine how the findings change when cyber-induced causes are considered. 
This two-stage approach, conventional analysis followed by cyber extension, is deliberate: it exposes the gap between what traditional safety methods reveal and what a cyber-aware analysis adds, and it demonstrates a practical pathway for adapting established safety methods to the cybersecurity domain.

The remainder of this paper is organized as follows. Section~\ref{sec:related-work} reviews related work on cyber risk assessment in critical infrastructure and hydropower systems. Section~\ref{sec:system-model} presents the system model and architecture of the smart hydropower plant, including the definition of the spillway gate control subsystem used as the shared analysis target. Section~\ref{sec:hazop-section} applies the HAZOP methodology and presents the cyber-extended deviation analysis.
Section~\ref{sec:bowtie-section} applies the BowTie method and examines how cyber threats interact with the barrier model. 
A comparative assessment of HAZOP and BowTie analysis is provided in Section~\ref{sec:comparitive_analysis}.
Advantages and limitations are presented in Section~\ref{sec:discussion}.
Finally, Section~\ref{sec:conclusion} concludes the paper and provides the future work. 


\section{Related Works}
\label{sec:related-work}

The growing dependence of critical infrastructure on networked control systems has prompted considerable work on cybersecurity risk assessment across several sectors. 
In the energy domain, 2015 cyber attacks on Ukrainian power distribution companies demonstrated that adversaries could achieve physical disruption through compromised industrial control systems~\cite{CISA2016Ukraine}. 
Analysis by Dragos subsequently identified the \texttt{CRASHOVERRIDE} malware used in a 2016 follow-up attack on Ukraine's transmission grid as the first malware purpose-built to disrupt electrical grid operations ~\cite{Dragos2017}. 
In 2021, the incident at the \texttt{Oldsmar}, Florida, water treatment facility raised sodium hydroxide levels to dangerous concentrations through a remotely accessible SCADA workstation, highlighting how vulnerable smaller utilities can be when control interfaces lack basic access controls.
The facility's systems shared a single password for remote access, running an unsupported operating system, and connected to the internet without firewall protection~\cite{CISA2021Oldsmar}.
These incidents have reinforced a broader recognition, articulated in early smart grid cybersecurity surveys~\cite{wang2013cyber}, that operational technology environments face threats that traditional safety analysis alone does not capture.

Several standards and frameworks have emerged in response. 
The IEC 62443 series provides a comprehensive framework for securing industrial automation and control systems, specifying security levels, zones, and conduits applicable across sectors~\cite{ISA62443}. 
NIST Special Publication 800-82 offers guidance specifically on securing SCADA, distributed control systems, and programmable logic controllers~\cite{NIST80082}. 
More recently, ISA-TR84.00.09 has addressed the integration of cybersecurity into the functional safety lifecycle, providing guidance on how cyber threats should be considered alongside traditional process hazards during hazard and risk analysis~\cite{ISATR84_2024}.
Within the hydropower sector, the U.S. Bureau of Reclamation's FIST 3-33 addresses ICS and SCADA operations and maintenance for hydropower facilities~\cite{USBR_FIST_3_33}, and the Pacific Northwest National Laboratory has developed guidance on cybersecurity response and recovery specifically for dam owners and operators~\cite{PNNL2020Recovery}.

On the methodological side, several established safety analysis techniques have been adapted or extended to incorporate security considerations. 
The concept of a cyber PHA or cyber HAZOP has gained traction in process industries, applying the familiar HAZOP deviation-based structure to cybersecurity threats against industrial control systems~\cite{ISATR84_2024}. 
Schmittner et al.~\cite{Schmittner2014} proposed extending the failure mode and effect analysis (FMEA) framework with a vulnerability cause-effect chain, creating \texttt{FMVEA}, a unified model for safety and security analysis. 
Young and Leveson~\cite{Young2014} approached the problem from a systems-theoretic perspective, arguing that STPA can be extended to security (STPA-Sec) by treating cyber vulnerabilities as inadequate control constraints rather than component failures. 
Abdo et al.~\cite{Abdo2018} combined BowTie analysis with an extended version of attack tree analysis to represent accidental and cyber-induced risk scenarios for industrial control systems, demonstrating the application on a chemical facility case study.
These works collectively illustrate a trend toward adapting traditional safety methods for the cyber domain, although specific adaptations vary considerably depending on the method and the application context.

In the maritime sector, where cyber-physical systems share structural similarities with hydropower control architectures, the application of risk assessment methods has progressed further. 
Sahay et al.~\cite{Sahay2023} applied STPA-Sec, STRIDE, and CORAS to a CyberShip framework, comparing how each method identifies threats and vulnerabilities when applied to the same system.
Their work found that each method has complementary strengths: STRIDE is effective at enumerating component-level threats, STPA-Sec identifies hazards arising from interactions between components, and CORAS provides visual risk modeling that supports communication with stakeholders.
They suggested that threats identified through STRIDE can serve as structured input to CORAS and STPA-Sec analyses, making them more systematic.
In~\cite{hyra2019analyzing}, the author used asset-based risk assessment methodology to identify cyber risks on major components of ship.

Within the hydropower sector specifically, cybersecurity research has been growing but remains less developed than in sectors such as oil and gas or maritime.
Sanghvi et al.~\cite{sanghvi2023hydropower} developed a value-at-risk framework for quantifying cybersecurity risk exposure in hydropower plants, specifying cyber risk in economic terms that can inform investment decisions.
The National Renewable Energy Laboratory has developed \texttt{CYSAT-Hydro}, a situational awareness tool designed to provide real-time monitoring of grid operations and network traffic for hydropower-integrated systems~\cite{NREL2024Hydro}. 
The Pacific Northwest National Laboratory has characterized the cyber-physical configurations typical of hydropower facilities~\cite{PNNL2021Hydro}, providing a reference architecture that supports more structured risk analysis. 
Alim et al.~\cite{alimlaboratory} developed a laboratory-scale spillway SCADA testbed for cybersecurity research, enabling controlled experimentation on the kinds of attack scenarios that would be too risky to test on operational infrastructure.

Despite this growing body of work, most hydropower cybersecurity studies have focused either on developing tools and frameworks at a relatively high level or on characterizing the attack surface in general terms.
The systematic application of established safety analysis methods, such as HAZOP and BowTie, to a hydropower control subsystem, with explicit extension to account for cyber-induced causes, has not been widely demonstrated. 
This paper identifies cyber and safety risks with HAZOP and BowTie in their conventional forms and then extending the analysis to examine how the findings change under cyber-induced causation.
The paper demonstrates a practical approach to adapting well-understood safety methods to the cybersecurity challenges facing networked hydropower control systems.

\section{System Architecture}
\label{sec:system-model}

\begin{figure}[h]
    \includegraphics[width=\textwidth]{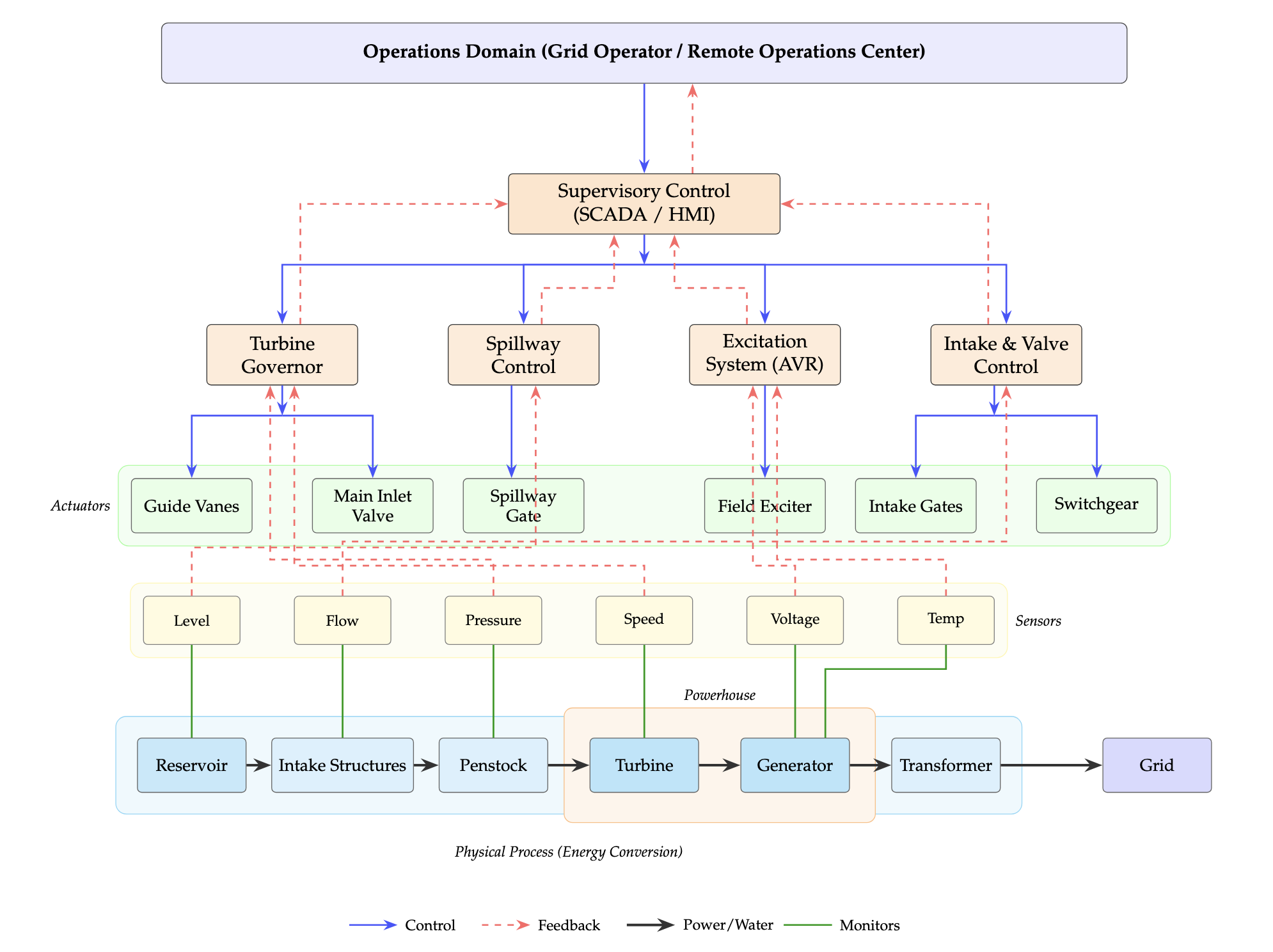}
    \centering
    \caption{ Hierarchical control structure of the smart hydropower plant.}
    \label{images:hydro-control-structure}
\end{figure}

The system architecture provides a representation of the physical power generation process, the associated instrumentation and control systems, and the interaction between infrastructure of a smart hydroelectric plant as a System under Consideration (SuC). 
It serves as a common reference architecture for the risk assessment methods applied in this paper.

Fig.~\ref{images:hydro-control-structure} provides a conceptual view of the hierarchical control structure of the systems. 
It shows the connection between operations domain, supervisory control, local control functions (e.g., Turbine Governor, Spillway Control, etc.), field instrumentation, i.e. actuators, and the physical energy conversion process. 
The components and signals shown are representative rather than exhaustive and are included to highlight interactions that are relevant to the system modeling and risk assessment performed in this study.

According to the NIST smart grid conceptual reference, the hydropower grid is also organized into seven interacting domains: customer, markets, service provider, operations, bulk generation, transmission, and distribution~\cite{NIST1108R1_2010}.
Although not all domains are relevant to the scope of this study.
Based on the system boundaries defined in our architectural analysis, this work focuses on a subset of the NIST domains that are directly involved in the hydro power grid, namely the generation, transmission, distribution, and operations domains. The remaining domains are treated as external to the SuC and, therefore, are not modeled in detail.

\subsection{Operational Workflow}
\label{subsec:grid-power-flow}

Within the defined scope, the smart hydroelectric plant operates as a bulk generation asset that injects electrical power into the transmission network.
From a physical perspective, electric power flows from the generation domain through the transmission infrastructure, is delivered to the distribution network, and ultimately supplies end users in the consumption domain. 
Although this sequence is conceptually simple, it provides essential context for understanding the role of the hydroelectric plant in maintaining grid stability and meeting demand.

In contrast to physical power transfer, the information required to operate and supervise the grid follows a more complex pattern.
As shown in Fig.~\ref{images:hydro-control-structure}, measurements, status information, and alarms originating from field sensors and local controllers are collected and processed by the supervisory control layer, which in turn communicates with the operations domain.
Control actions and setpoints flow downward from the operations domain through the supervisory layer to local controllers and actuators, regulating generation output, protecting equipment, and responding to abnormal conditions (based on the conceptual framework in~\cite{NIST1108R1_2010}).

\subsection{Physical Process Components}
\label{subsec:physical-components}

A hydroelectric plant converts the potential energy of stored water into electricity through a series of interconnected physical subsystems. The following component descriptions are informed by established hydroelectric engineering references \cite{IITR_Hydro_Control, PNNL2021Hydro} and the IEEE standards for turbine governing and plant control \cite{IEEE1207_2011, IEEE1010_2022}.

\begin{enumerate}
    \item \textbf{Reservoir:} A natural or artificial body of water that stores the plant's energy source. The vertical distance between the reservoir surface and the turbine, known as the hydraulic head, determines the available pressure for generation.
    
    \item \textbf{Intake Structures:} Positioned where water enters the plant. Screens and trash racks prevent debris from damaging downstream equipment.

    \item \textbf{Penstock:} A large pipe or conduit that carries pressurized water from the intake to the turbine.
    
    \item \textbf{Turbine:} A rotating machine that extracts energy from moving water. Water enters through adjustable openings called guide vanes or wicket gates, which control flow rate and consequently the turbine's speed and power output.
    
    \item \textbf{Generator:} Coupled with the turbine shaft, the generator converts mechanical rotation into electrical power.
    
    \item \textbf{Transformers:} Increase the generator's output voltage to the higher levels required for long-distance transmission.
    
    \item \textbf{Powerhouse:} The structural enclosure that houses the turbine and generator, as shown in Fig.~\ref{images:hydro-control-structure}.
    It provides mechanical support, environmental protection, and access for operation and maintenance, and serves as the central location where hydraulic energy is converted into electrical energy.

    \item \textbf{Grid Interface:} The point at which the plant's electrical output connects to the external transmission network. 
    The grid receives the increased voltage from the transformers and represents the boundary between the plant and the wider power system.
\end{enumerate}

\subsection{Controllers and Actuators}
\label{subsec:controllers-actuators}

The physical process components are regulated and monitored by controllers and actuators that together enable safe and responsive plant operation. These components form the interface between the digital control systems and the physical process and are the elements most directly relevant to the risk analysis performed in later sections.

\paragraph{Controllers}
\begin{itemize}[nosep]
    \item \textbf{Turbine Governor:} Automatically adjusts the turbine's guide vanes or wicket gates to maintain stable turbine speed and respond to changing load demands.
    \item \textbf{Spillway Control:} Manages spillway gate position based on reservoir level measurements to maintain safe water levels.
    \item \textbf{Excitation System (AVR):} Controls the magnetic field strength inside the generator by supplying regulated current to the field winding, thereby regulating output voltage.
    \item \textbf{Intake and Valve Control:} Manages the positions of the intake gates and switchgear based on flow measurements and operational commands.
\end{itemize}

\paragraph{Actuators}
Actuators convert control commands into physical actions on the process:
\begin{itemize}[nosep]
    \item \textbf{Guide Vanes:} Adjustable openings at the turbine inlet that regulate water flow rate under command from the turbine governor.
    \item \textbf{Main Inlet Valve:} Located upstream of the turbine, this valve allows operators to rapidly shut off flow for emergencies or maintenance.
    It serves as a critical isolation point between hydraulic and mechanical subsystems.
    \item \textbf{Spillway Gate:} Provides controlled discharge of excess water when reservoir levels exceed safe limits or during flood conditions.
    \item \textbf{Field Exciter:} Supplies regulated current to the generator's field winding under command from the excitation system, controlling the generator's magnetic field strength and output voltage.
    \item \textbf{Intake Gates:} Control the volume of water admitted from the reservoir into the penstock, regulating flow according to generation requirements.
    \item \textbf{Switchgear:} Circuit breakers and disconnect switches that provide protection and allow sections of the electrical system to be isolated for safety or maintenance.
\end{itemize}

\paragraph{Field Instrumentation}
Sensors provide the real-time measurements that close the feedback loops on which automatic control depends. 
Typical measurements include reservoir level, water flow and pressure, turbine speed, generator voltage, and temperatures of critical components such as generator windings and bearings. 
These measurements form the primary input to the controllers and supervisory systems described in Section~\ref{subsec:instrumentation-control-communication}.

\subsection{Hydropower Conversion Process}
\label{subsec:hydro-power-conversion}

The physical components and their associated controllers and actuators introduced in the preceding subsections interact across a sequence of functional stages to convert the potential energy of stored water into electrical energy suitable for grid injection. 
This functional decomposition is commonly used in hydroelectric engineering and provides a clear basis for analyzing plant operation and control~\cite{IITR_Hydro_Control}.

\paragraph{Stage 1: Water Storage, Intake Regulation, and Spillway Management}
The process begins at the reservoir, which provides both hydraulic head and operational flexibility for generation scheduling.
The intake gates regulate the volume of water admitted according to the generation requirements and coordinate the plant output with the system-level dispatch signals. 
Spillway gates provide a controlled discharge path for excess water when reservoir levels approach or exceed safe limits, such as during flood conditions or periods of low generation demand.
Proper coordination of intake and spillway operations is essential to maintain reservoir levels within safe operating limits while meeting generation targets.

\paragraph{Stage 2: Water Conveyance and Isolation}
From the intake, water flows through the penstock under pressure toward the powerhouse. 
The main inlet valve, located upstream of the turbine, serves as a critical isolation point between the hydraulic and mechanical subsystems.
This valve enables rapid shutdown during abnormal operating conditions, protects the turbine during maintenance activities, and provides a safety barrier against uncontrolled water entry. 
The penstock and inlet valve together define an important operational boundary within the plant.

\paragraph{Stage 3: Mechanical Energy Conversion and Governing}
Mechanical energy conversion occurs in the turbine, where pressurized water acts on the runner to produce rotational motion. 
The \texttt{Turbine Governor} continuously monitors the turbine speed and adjusts the guide vane or wicket gate to regulate flow and maintain stable operation. 
This closed-loop control enables the turbine to respond to load changes, frequency deviations, and dispatch commands while avoiding over-speed or unstable operating conditions~\cite{IEEE1207_2011, UNIDO_SHP_TG003_3}.

\paragraph{Stage 4: Electrical Energy Conversion and Voltage Regulation}
The rotating turbine shaft drives the generator, converting mechanical energy into three-phase electrical power.
The excitation system controls the generator's magnetic field strength to regulate output voltage and support reactive power requirements.
An automatic voltage regulator (AVR) provides closed-loop control to maintain voltage within specified limits under varying load conditions. The supporting subsystems, including the cooling, lubrication, and bearing systems, support the continuous and reliable operation of the rotating equipment~\cite{IITR_Hydro_Control,IEEE1010_2022}.

\paragraph{Stage 5: Grid Interface and Protection}
The generated power is conditioned and delivered to the transmission network through the plant's electrical infrastructure.
Step-up transformers raise the generator output voltage to transmission levels, reducing losses over long distances.
Switchgear, including circuit breakers and disconnect switches, provides the ability to isolate the generator from the grid during faults, maintenance, or abnormal system conditions. 
Protection relays monitor electrical parameters and initiate tripping sequences when predefined thresholds are exceeded.
This stage defines the electrical boundary between the hydroelectric plant and the external grid and is the point at which the plant's operation interfaces with broader requirements of the power system.

The plant's monitoring and supervisory functions are typically supported by industrial control systems, including SCADA components, which require defined operational and maintenance practices~\cite{USBR_FIST_3_33}.

\subsection{Instrumentation, Control, and Communication Architecture}
\label{subsec:instrumentation-control-communication}

The controllers, actuators, and sensors described in Section~\ref{subsec:controllers-actuators} are implemented through programmable logic controllers (PLCs) and dedicated control units that execute control logic based on sensor inputs and predefined setpoints. 
Table~\ref{tab:control-loops} summarizes the main control loops, mapping each controller to its associated actuators, sensors, and primary function.

\begin{table}[ht]
\centering
\caption{Principal control loops in the smart hydropower plant.}
\label{tab:control-loops}
\footnotesize
\begin{tabular}{>{\raggedright\arraybackslash}p{3cm}
                >{\raggedright\arraybackslash}p{3.5cm}
                >{\raggedright\arraybackslash}p{2.5cm}
                >{\raggedright\arraybackslash}p{3.5cm}}
\hline
\textbf{Controller} & \textbf{Actuators} & \textbf{Sensors} & \textbf{Primary Function} \\
\hline
Turbine Governor & Guide Vanes, Main Inlet Valve & Speed, Pressure & Speed/power regulation \\
Spillway Control & Spillway Gate & Level & Reservoir level management \\
Excitation System & Field Exciter & Voltage, Temp & Voltage regulation \\
Intake \& Valve Control & Intake Gates, Switchgear & Flow & Water admission, grid isolation \\
\hline
\end{tabular}
\end{table}

Each control loop operates as a closed-loop system, receiving sensor feedback and adjusting actuator output to maintain desired operating conditions. 
The integrity of these feedback paths is essential for safe and stable operation; corruption or loss of sensor data can lead to incorrect control actions, oscillatory behavior, or failure to respond to abnormal conditions.

\paragraph{Supervisory Control and Monitoring}
Supervisory control and data acquisition (SCADA) systems provide higher-level monitoring, control, and data logging capabilities. 
As shown in the upper levels of Fig.~\ref{images:hydro-control-structure}, SCADA systems collect measurements and status information from local controllers, present the status of the plant to operators through human-machine interfaces (HMI), and allow supervisory commands and setpoints to be issued. 
In hydroelectric plants, the supervisory functions support both the local control room and remote monitoring from centralized operations centers~\cite{USBR_FIST_3_33}.

\paragraph{Communication Infrastructure}
Communication networks interconnect sensors, controllers, and supervisory systems within the plant and link the plant to external operational domains.
These networks support the exchange of measurements, alarms, status information, and control commands.
As shown in Fig.~\ref{images:hydro-control-structure}, vertical communication paths represent information flows between hierarchical levels.
Standardized communication architectures and data models are increasingly being used to improve the interoperability and integration of hydroelectric plants within smart grid environments.

\paragraph{Operational Boundaries.}
The instrumentation and control architecture defines the operational boundaries between field devices, local control systems, and supervisory functions, as reflected in the layered structure of Fig.~\ref{images:hydro-control-structure}. 
These boundaries support safe operation, maintenance activities, and system integration.
They also form an important foundation for the analysis of threat scenarios and possible attack paths, as compromises at different levels of the hierarchy can have different consequences for plant safety and operation~\cite{USBR_FIST_3_33}.

\subsection{System under Consideration (SuC)}
\label{sec:suc}

To perform a risk assessment, SuC must be identified~\cite{ISA62443}.
Clearly defined SuC sets the boundary for the assessment because organizations have many control systems and field devices.
For the cyber and safety assessment SuC is the Spillway Gate control.

\subsection{Analysis Subsystem: Spillway Gate Control}
\label{subsec:analysis-subsystem}

The risk assessment methods applied in the following sections require a concrete subsystem to serve as the analysis target. 
This subsection specifies the selected subsystem and its boundary.

The spillway gate control subsystem is selected as the focus for detailed risk analysis throughout this work. 
This selection is motivated by three considerations. 
First, spillway gates are critical safety components of dams whose reliability directly impacts the assessment of dam risk and operational performance~\cite{ballard2004reliability}.
Second, studies in the hydroelectric sector identify gates and water conveyance systems as cyber-dependent assets that warrant explicit consideration in cyber-physical risk assessments~\cite{sanghvi2023hydropower}.
Third, the subsystem shows the combination of physical actuation, digital control, sensor feedback, and operator interaction that characterizes the broader plant architecture, making it a representative case for examining both conventional and cyber-induced hazards.

\begin{figure}[h]
    \includegraphics[width=1.0\linewidth, height=2.0\textheight, keepaspectratio]{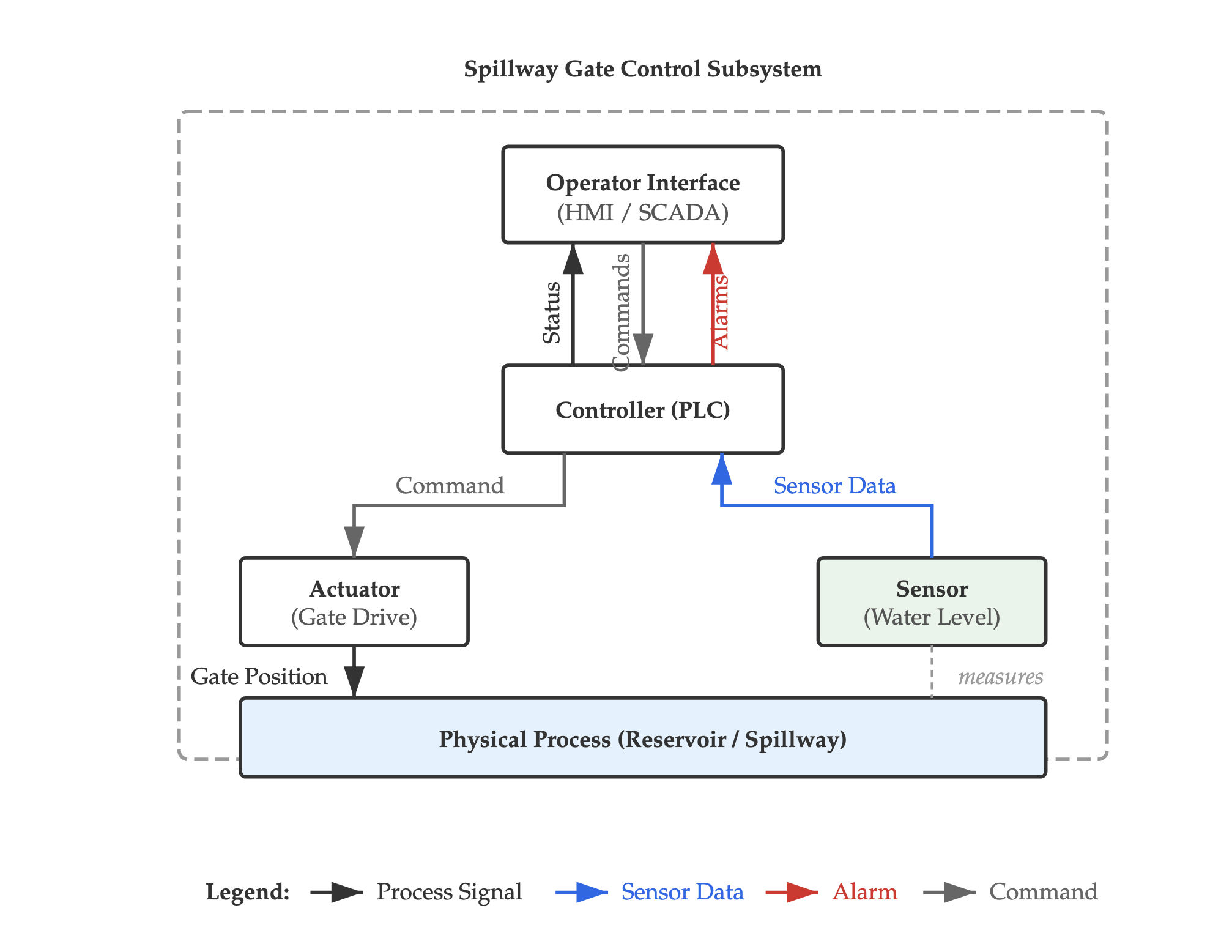}
    \centering
    \caption{Functional control structure of the spillway gate control subsystem}
    \label{images:spillway_control_structure}
\end{figure}

Fig.~\ref{images:spillway_control_structure} highlights the functional control structure of the spillway gate control subsystem.
This representation aligns with documented SCADA testbed architectures for spillways in which water level sensors interface with a programmable logic controller (PLC), which connects to an operator interface for supervisory monitoring and control via supporting communication links~\cite{alimlaboratory}. 
The structure comprises five principal elements arranged in a closed control loop:

\begin{itemize}[nosep]
    \item \textbf{Operator Interface (HMI/SCADA):} It provides the human-machine interface through which operators monitor reservoir status, receive alarms, and issue setpoints or manual commands to the controller.
    \item \textbf{Controller (PLC):} Executes the control logic, comparing measured water level with setpoints to determine appropriate gate positions.
    The controller issues commands to the actuator and transmits status and alarm information to the operator interface.
    \item \textbf{Actuator (Gate Drive):} Receives control commands from the Programmable Logic Controllers (PLCs) and converts them into physical gate movement, adjusting the spillway gate position to regulate discharge flow.
    \item \textbf{Sensor (Water Level):} Measures the reservoir water level and transmits this measurement to the controller, closing the feedback loop that enables automatic level regulation.
    \item \textbf{Physical Process (Reservoir/Spillway):} The controlled process itself, where water accumulates in the reservoir from upstream inflows and is discharged through the spillway gates based on gate position.
\end{itemize}

The analysis boundary includes the physical gate mechanism, the sensing and actuation elements, the control logic, the operator interaction, and the communication paths required to support these functions. 
External systems lie outside the analysis boundary and are not modeled in detail, except where their interfaces directly influence the control behavior of the spillway gate (e.g., supervisory setpoints or remote monitoring paths). 
This boundary definition follows the cyber-physical configuration framework established for hydroelectric systems~\cite{PNNL2021Hydro}.

This definition of the subsystem provides the shared analysis target for the HAZOP study in Section~\ref{sec:hazop-analysis}, the BowTie analysis in Section~\ref{sec:bowtie-analysis}, and the cyber-extended assessments that follow.

\section{Hydropower Plant Analysis using HAZOP}
\label{sec:hazop-section}

The spillway gate control subsystem defined in Section~\ref{subsec:analysis-subsystem} is the target for the HAZOP study that follows.

\subsection{HAZOP Methodology and Application}
\label{sec:hazop-analysis}

Hazard and Operability Study (HAZOP) is a systematic technique for identifying hazards and operability issues in process systems.
The method originated at Imperial Chemical Industries (ICI) in the United Kingdom during the 1960s and was formalized following major industrial accidents that demonstrated the need for structured hazard identification~\cite{Crawley2015, Kletz1999}. 
HAZOP is founded on the principle that hazards arise when process conditions deviate from their intended design parameters.
A multidisciplinary team systematically examines each part of a system applying standardized \textit{guide words}, such as NO, MORE, LESS, REVERSE, and OTHER THAN, to relevant process parameters, generating potential deviations whose causes, consequences, and safeguards are then analyzed~\cite{IEC61882}.

Although HAZOP was developed for chemical process plants, its systematic deviation-based approach has been adapted for control systems and other domains~\cite{Elhosary2024, Choi2020}.
The node-based structure naturally maps to functional subsystems and control loops. 
The guide words can express deviations in sensor readings, control signals, and actuator responses. 
The cause-consequence-safeguard framework accommodates both hardware failures and software or logic errors. 
These characteristics make HAZOP applicable to the analysis of control subsystems within the hydroelectric plant system model presented in Section~\ref{sec:system-model}.

\subsection{Analysis Node and Parameters}

The spillway gate control subsystem (Section~\ref{subsec:analysis-subsystem}) serves a critical safety function: releasing excess water from the reservoir to prevent overtopping of the dam during high inflow conditions or to maintain reservoir levels within operational limits.
Failure to operate spillway gates correctly, whether due to equipment malfunction, operator error, or malicious interference, can result in consequences ranging from downstream flooding to structural damage to the dam itself.

In HAZOP analysis, the control loop described in Fig.~\ref{images:spillway_control_structure} is considered as a single analysis node. 
The key parameters subject to deviation correspond to the signals and states flowing through the control structure:

\begin{itemize}[nosep]
    \item \textbf{Water Level (measured):} The sensor reading provided to the controller.
    \item \textbf{Setpoint:} The target water level or gate position commanded by the operator.
    \item \textbf{Control Command:} The signal from the controller to the actuator that specifies the position of the gate. 
    This determines the direction of gate movement (open or closed) and the extent of opening, which directly controls the discharge flow rate. 
    \item \textbf{Gate Position:} The actual physical position of the 
    spillway gate, indicating how far the gate is open or closed. 
    \item \textbf{Status/Alarms:} Information reported from controller to operator interface such as alarms and water level.
\end{itemize}

\subsection{Deviation Analysis}

For each parameter, the standard HAZOP guide words are applied to generate deviations. 
The causes, consequences, and existing safeguards for each credible deviation are then analyzed. 
Table~\ref{tab:hazop-spillway} presents the results of this analysis.

\begin{sidewaystable}
\centering
\setlength{\tabcolsep}{4pt}
\caption{HAZOP Analysis of Spillway Gate Control Subsystem}
\label{tab:hazop-spillway}
\footnotesize
\begin{tabular}{>{\raggedright\arraybackslash}p{1.6cm}
                >{\raggedright\arraybackslash}p{1.4cm}
                >{\raggedright\arraybackslash}p{2.4cm}
                >{\raggedright\arraybackslash}p{3.2cm}
                >{\raggedright\arraybackslash}p{3.8cm}
                >{\raggedright\arraybackslash}p{5.2cm}}
\hline
\textbf{Parameter} & \textbf{Guide Word} & \textbf{Deviation} & \textbf{Possible Causes} & \textbf{Consequences} & \textbf{Existing Safeguards} \\
\hline

Water Level (measured) & NO & No level signal received by controller & Sensor failure; cable fault; power loss to sensor & Controller loses feedback; may fail to open-state or hold last position; potential overflow or underflow & Redundant level sensors; signal loss alarm; watchdog timer on sensor input \\
 & MORE & Level reading higher than actual & Sensor drift; calibration error; electrical interference & Gates open unnecessarily; excessive discharge; downstream impact; wasted water & Periodic sensor calibration; plausibility checks against secondary measurements \\
 & LESS & Level reading lower than actual & Sensor fouling; zero drift; damaged sensor element & Gates remain closed when opening required; reservoir overtopping risk & Regular sensor maintenance; high-level backup alarm; manual inspection procedures \\
 & OTHER THAN & Spurious or corrupted signal & EMI; loose connection; sensor malfunction & Erratic gate operation; nuisance alarms; operator confusion & Signal filtering; shielded cabling; out-of-range detection \\
\hline
Setpoint & NO & No setpoint provided & HMI failure; communication loss; operator absent & Controller uses default or last setpoint; may not respond to changing conditions & Default safe setpoint; communication heartbeat monitoring; operator presence checks \\
 & MORE & Setpoint higher than intended & Operator input error; HMI display fault & Water level rises above safe limit; potential overtopping & Setpoint limit validation; confirmation prompts for large changes; high-level alarms \\
 & LESS & Setpoint lower than intended & Operator input error; unit conversion mistake & Excessive discharge; downstream flooding; reduced generation capacity & Minimum setpoint limits; downstream flow alarms; operator training \\
 & OTHER THAN & Wrong parameter modified & HMI design flaw; operator selects wrong control & Unintended system behaviour; potential cascade effects & Clear HMI labelling; role-based access control; change logging \\
\hline
Control Command & NO & No command reaches actuator & PLC failure; network fault; output module failure & Gate remains in last position; loss of automatic control & Redundant communication paths; PLC watchdog; manual override capability \\
 & MORE & Command requests excessive opening & Logic error; setpoint spike; calculation overflow & Rapid gate opening; downstream surge; hydraulic transients & Rate-of-change limits on gate movement; command validation logic \\
 & LESS & Command requests insufficient opening & Logic error; output saturation; signal attenuation & Inadequate discharge; rising reservoir level & Position feedback verification; high-level alarms independent of control loop \\
 & REVERSE & Command direction inverted & Wiring error; software bug; configuration mistake & Gate moves opposite to intended; dangerous condition & Factory acceptance testing; commissioning verification; position feedback monitoring \\
\hline
Gate Position & NO & Gate does not move & Mechanical jam; actuator failure; power loss & Loss of discharge control; reservoir level uncontrolled & Position feedback monitoring; actuator health diagnostics; backup power supply \\
 & PART OF & Gate partially responds & Obstruction; actuator degradation; hydraulic leak & Reduced discharge capacity; unpredictable control & Travel time monitoring; periodic stroke testing; maintenance schedules \\
 & MORE & Gate opens further than commanded & Limit switch failure; actuator runaway; brake failure & Excessive discharge; potential structural stress & Hard-wired limit switches; mechanical stops; position discrepancy alarm \\
 & LESS & Gate opens less than commanded & Friction; ice formation; debris; wear & Insufficient discharge; level rises & Command vs.\ feedback comparison; maintenance inspection; debris screens \\
\hline
Status / Alarms & NO & No status reported to operator & Communication failure; HMI fault; display frozen & Operator unaware of system state; delayed response to abnormal conditions & Communication heartbeat; HMI watchdog; independent local indicators \\
 & OTHER THAN & False or misleading status & Software bug; display mapping error; stale data & Operator takes incorrect action based on wrong information & Data validation checks; timestamp display; periodic HMI testing \\
\hline
\end{tabular}
\end{sidewaystable}

The analysis in Table~\ref{tab:hazop-spillway} identifies deviations arising from traditional causes: equipment failures, sensor malfunctions, communication losses, wiring errors, and human operating errors. 
These represent the types of hazards that traditional HAZOP is designed to uncover, and the safeguards identified would typically inform engineering measures such as redundant sensors, watchdog timers, alarm management systems, and operator training.

\subsection{Cyber Considerations in the HAZOP Framework}

The traditional HAZOP analysis in Table~\ref{tab:hazop-spillway} identifies deviations whose causes are accidental: sensor drift, cable faults, operator input errors, etc.
These causes are generally independent of one another, and the safeguards identified, such as redundant sensors, watchdog timers, and alarm management, are designed with that assumption. 
They expect failures to be random and uncoordinated.

A cyber attacker operating within the control network does not face such constraints.
An adversary who gains access to the sensor communication path could inject false water level readings, causing the controller to perceive conditions that do not exist.
This class of attack, broadly categorized as false data injection in the smart grid security literature~\cite{wang2013cyber}, has been extensively studied in the context of power system state estimation, but its application to hydropower control loops has received less attention. 
The same attacker could simultaneously suppress the alarms that would normally alert operators to the discrepancy. 
Where a conventional sensor failure triggers a signal loss alarm and is quickly noticed, a well-crafted cyber attack can produce a reading that looks entirely normal to both the controller and the operator, yet bears no relation to the actual reservoir state.

Table~\ref{tab:cyber-hazop-demo} re-examines two deviations from the conventional HAZOP under cyber-induced causation, replacing accidental causes with attack scenarios, and identifying the security-specific safeguards that conventional measures do not cover.

\begin{sidewaystable}
\centering
\setlength{\tabcolsep}{4pt}
\caption{Cyber-Extended HAZOP: Representative Deviations with Cyber-Induced Causes}
\label{tab:cyber-hazop-demo}
\footnotesize
\begin{tabular}{>{\raggedright\arraybackslash}p{1.6cm}
                >{\raggedright\arraybackslash}p{1.2cm}
                >{\raggedright\arraybackslash}p{2.2cm}
                >{\raggedright\arraybackslash}p{3.5cm}
                >{\raggedright\arraybackslash}p{4.0cm}
                >{\raggedright\arraybackslash}p{5.0cm}}
\hline
\textbf{Parameter} & \textbf{Guide Word} & \textbf{Deviation} & \textbf{Cyber-Induced Causes} & \textbf{Additional Consequences} & \textbf{Security-Specific Safeguards} \\
\hline
Water Level (measured) & LESS & Level reading lower than actual & Man-in-the-middle attack on sensor communication link, injecting artificially low readings; compromised PLC firmware reporting manipulated values; replay attack substituting previously recorded low-level data & Controller keeps gates closed during rising water; reservoir overtopping risk, as with conventional cause, but attack may be timed to coincide with high inflow for maximum effect; conventional alarm may also be suppressed if attacker has network access, delaying operator awareness & Authenticated sensor communication (e.g., message authentication codes on sensor data); network traffic anomaly detection on sensor links; independent out-of-band level verification (e.g., physical gauge or secondary sensor on separate network); firmware integrity monitoring on PLC; periodic comparison of sensor readings against independent sources \\
\hline
Status / Alarms & NO & No status reported to operator & Denial-of-service attack on HMI communication path; selective packet filtering to block alarm traffic while allowing control traffic to pass; compromise of HMI application to suppress alarm display; modification of alarm thresholds in controller configuration to prevent alarm generation & Operator remains unaware of developing hazard; if combined with sensor manipulation (see row above), operator has no independent indication that conditions are abnormal; response time to incident is extended; post-incident investigation is hampered by missing records & Network segmentation between alarm and control traffic paths; independent alarm annunciation system not connected to primary control network; alarm heartbeat monitoring (absence of any alarm traffic within expected interval triggers a secondary alert); write protection on alarm configuration parameters; audit logging on separate, append-only storage \\
\hline
\end{tabular}
\end{sidewaystable}

The two rows in Table~\ref{tab:cyber-hazop-demo} represent a plausible coordinated attack: an adversary suppresses the alarm channel (second row) while simultaneously feeding the controller a false low water level reading (first row). 
The controller, seeing no reason to open the gates, holds them closed. The operator, receiving no alarms, has no reason to intervene.
Meanwhile, the actual reservoir level increases.
Neither deviation in isolation is unusual from a safety perspective; both appear in the conventional HAZOP.
What makes the cyber case qualitatively different is the coordination between them and the deliberate timing to maximize impact.

This example also highlights a gap in the traditional safeguard set.
The traditional HAZOP identifies ``redundant level sensors'' as a safeguard against loss of level signal.
However, if both sensors communicate over the same network segment, a single network-level attack could compromise both. 
The security-specific safeguards in Table~\ref{tab:cyber-hazop-demo} address this by requiring physical separation of communication paths and independent verification mechanisms that do not rely on the same infrastructure as the primary control loop.

The pattern illustrated by these two deviations is consistent across the entire set in Table~\ref{tab:hazop-spillway}: cyber-induced causes tend to be more targeted than accidental ones, consequences can be amplified through coordination, and the safeguards required go beyond redundancy and into authentication, segmentation, and independent verification.



\section{Hydropower Plant Analysis using BowTie}
\label{sec:bowtie-section}

The BowTie method provides a different perspective on the same subsystem.
While HAZOP systematically generates deviations and their causes, BowTie maps the relationship between threats, a central unwanted event, and the barriers that stand between them.

\subsection{BowTie Methodology and Application}
\label{sec:bowtie-analysis}

The BowTie method is a structured risk assessment technique that visualizes the relationship between threats, an unwanted event, and potential consequences, together with barriers designed to prevent or mitigate harm. 
The method emerged from the investigation of major industrial accidents, where traditional risk tools were found to be insufficient to communicate how incidents develop and spread~\cite{DeRuijter2016}.
Since its formalization, BowTie analysis has been widely adopted in high-hazard industries including oil and gas, aviation, and chemical processing, where it supports both risk assessment and operational decision-making~\cite{CCPS2018}.

The structure of a BowTie diagram places an unwanted event, often termed the \textit{top event}, at the center. 
The left side of the diagram identifies the threats or causes that could lead to the event, while the right side maps the consequences that could result if the event occurs. 
Barriers are positioned on both sides: \textit{preventive barriers} (or controls) on the left aim to stop threats from escalating into the top event, while \textit{mitigative barriers} (or recovery measures) on the right aim to limit the severity of consequences once the event has occurred.
This dual-sided structure offers a complete view of both proactive and reactive risk controls~\cite{CCPS2018, Ferdous2013}.

A key strength of the BowTie method is its ability to represent complex risk scenarios in a visual format that is accessible to engineers, operators, and management alike.
BowTie diagrams communicate the essential risk logic in a single integrated view. 
This makes them particularly valuable for safety communication, training, and management review~\cite{DeRuijter2016}. 
The method also supports the identification of \textit{escalation factors}, conditions that can degrade or defeat barriers, helping to reveal vulnerabilities in the overall risk control framework.

BowTie analysis is well suited to cyber-physical systems such as hydroelectric plants, where a single unwanted event can be triggered by physical faults, operational errors, or cyber intrusions, and where these causes may interact. 
Similarly, barriers can include technical controls (such as network segmentation or emergency shutdown systems), procedural measures (such as verification steps or emergency protocols), and human interventions (such as manual overrides), reflecting the layered defenses typical of industrial control systems.

The continuity with the preceding HAZOP study allows us to demonstrate how the HAZOP findings feed directly into the BowTie analysis: deviations identified through HAZOP become threats on the left side of the BowTie, while the causes, consequences, and safeguards documented in the HAZOP worksheet inform the barrier identification process.

\subsection{Top Event Definition}

The top event for this BowTie analysis is \textbf{loss of spillway gate control}: the condition in which the ability to command or regulate spillway gate position is compromised, either due to equipment failure, loss of communication, incorrect control signals, or deliberate interference. 
This event indicates the boundary between normal operation and a potentially hazardous state; at this point, the system can no longer reliably regulate reservoir discharge, but the full consequences have not yet materialized.

This event corresponds directly to the HAZOP deviations identified in Section~\ref{sec:hazop-analysis} (such as sensor signal loss, incorrect control commands, or actuator failure) and provides the central node from which threats and consequences are mapped in the BowTie structure that follows.

\subsection{Threats and Preventive Barriers}

The left side of the BowTie identifies threats that could lead to loss of control of the spillway gate. 
Drawing on the HAZOP analysis and the cyber-physical nature of the system, these threats can be grouped into three categories:

\textbf{Cyber threats} include scenarios in which digital systems are compromised or manipulated. 
Spoofed or corrupted sensor data could cause the controller to perceive incorrect reservoir levels, leading to inappropriate gate commands.
Blocked or delayed operator commands could prevent timely gate adjustments during abnormal conditions. 
Unauthorized remote access could enable an adversary to issue malicious commands directly.
Malware infection in the control system could disrupt the execution of normal logic or disable safety functions.

\textbf{Physical threats} encompass conventional equipment failures. Hydraulic system failures (such as pump failure, fluid leaks, or valve faults) could prevent the actuator from moving the gate.
Mechanical obstructions (such as debris, ice, or structural deformation) could physically hinder the movement of the gates. 
Electrical supply failures could disable the controller, actuator, or sensors. 
Sensor failures (such as drift, fouling, or complete loss of signal) could deprive the controller of the feedback needed for regulation.

\textbf{Operational threats} arise from human actions or procedural failures. 
Misconfigured control logic (such as incorrect setpoints, faulty tuning, or software errors) could cause the controller to issue inappropriate commands.
Delayed or incorrect operator response could allow abnormal conditions to escalate. 
Inadequate maintenance could leave the equipment in a degraded state, increasing the chances of failure under stress.

Each of these threats is addressed by one or more \textit{preventive barriers} designed to stop the threat from escalating into the top event.
Table~\ref{tab:bowtie-preventive} summarizes the key preventive barriers identified for each category of threat.

\begin{table}[ht]
\centering
\caption{Preventive Barriers for Loss of Spillway Gate Control}
\label{tab:bowtie-preventive}
\footnotesize
\begin{tabular}{>{\raggedright\arraybackslash}p{2.8cm}
                >{\raggedright\arraybackslash}p{4.5cm}
                >{\raggedright\arraybackslash}p{5.5cm}}
\hline
\textbf{Threat Category} & \textbf{Example Threats} & \textbf{Preventive Barriers} \\
\hline
Cyber & Spoofed sensor data; blocked commands; unauthorized access; malware & Network segmentation; authentication and access control; input validation and plausibility checks; intrusion detection; secure remote access protocols; application whitelisting \\
\hline
Physical & Hydraulic failure; mechanical obstruction; power loss; sensor failure & Redundant sensors; backup power supply (UPS/generator); mechanical interlocks; regular inspection and maintenance; environmental protection (debris screens, heating) \\
\hline
Operational & Misconfigured logic; operator error; delayed response; inadequate maintenance & Configuration management and change control; operator training and competency verification; alarm management; SCADA validation checks; maintenance scheduling and tracking \\
\hline
\end{tabular}
\end{table}

\subsection{Consequences and Mitigative Barriers}

The right side of the BowTie identifies the consequences that could occur if loss of spillway gate control occurs and is not contained. 
The severity of consequences depends on the reservoir state, inflow conditions, duration of the loss of control, and effectiveness of mitigative measures.

\textbf{Uncontrolled water release} could occur if gates fail in an open position or open unexpectedly, resulting in excessive discharge downstream. 
This could cause flooding, erosion, damage to downstream infrastructure, or harm to people in the flood zone.

\textbf{Reservoir instability} could occur if gates fail to open when required during high inflow conditions, causing reservoir levels to rise beyond safe limits. 
In extreme cases, this could lead to dam overtopping, with potential for structural damage or catastrophic failure.

\textbf{Turbine and generator impacts} could arise from rapid changes in water flow affecting the power generation equipment. 
Sudden loss of head or flow could cause turbine over speed, cavitation, or mechanical stress, potentially triggering protective trips or causing equipment damage.

\textbf{Structural stress} on the dam or spillway structure could result from abnormal water levels, unbalanced loading, or hydraulic transients caused by erratic gate operation.

These consequences are addressed by \textit{mitigative barriers} designed to limit harm once the top event has occurred. Table~\ref{tab:bowtie-mitigative} summarizes the key mitigation barriers.

\begin{table*}[ht]
\centering
\caption{Mitigative Barriers Following Loss of Spillway Gate Control}
\label{tab:bowtie-mitigative}
\footnotesize
\setlength{\tabcolsep}{8pt}
\renewcommand{\arraystretch}{1.8}
\begin{tabular}{>{\raggedright\arraybackslash}p{3.5cm}
                >{\raggedright\arraybackslash}p{12cm}}
\hline
\textbf{Barrier Type} & \textbf{Mitigative Barriers} \\
\hline
Technical / Automated & Emergency gate closure mechanism (failsafe); automated high-level shutdown logic; protective relays for turbine/generator; independent backup gate or auxiliary spillway \\
\hline
Monitoring / Warning & Independent high-water-level alarms; downstream flood warning systems; real-time monitoring dashboards; out-of-band status verification \\
\hline
Procedural / Human & Manual override capability (local control); emergency operating procedures; operator training for abnormal conditions; coordination with downstream authorities and emergency services \\
\hline
\end{tabular}
\end{table*}

\subsection{Escalation Factors}

An important element of the BowTie analysis is the identification of \textit{escalation factors}, conditions that can reduce the effectiveness of preventive or mitigative barriers. 
In a cyber-physical context, escalation factors deserve particular attention because cyber attacks can specifically target barrier systems.

\textbf{Cyber-induced escalation} could occur if an attacker disables alarms, corrupts monitoring displays, or prevents communication between the control system and operators. 
For example, an adversary who compromises the intrusion detection system (a preventive barrier against cyber threats in Table~\ref{tab:bowtie-preventive}) while simultaneously injecting false sensor data can defeat two independent barriers in a coordinated manner, significantly increasing the likelihood that the top event leads to severe consequences. 
More broadly, an attack that compromises both the primary control path and the safety monitoring systems would reduce the effectiveness of mitigative barriers that depend on timely operator awareness.

\textbf{Environmental escalation} could arise from extreme weather, flooding, or seismic events that stress multiple systems simultaneously, potentially causing common-cause failures across redundant barriers.

\textbf{Human and organizational escalation} could result from inadequate training, fatigue, or poor communication during an incident, reducing the effectiveness of procedural barriers. 
Maintenance backlogs or deferred inspections could leave physical barriers in a degraded state.

\textbf{Communication failures} could delay notification of abnormal conditions, slow the activation of the emergency response, or prevent coordination with external agencies.

Identifying these escalation factors is important as it highlights where additional controls or redundancies may be needed to maintain barrier integrity under adverse conditions.

\subsection{BowTie Diagram for Loss of Spillway Gate Control}

\begin{figure}[h]
    \includegraphics[width=\textwidth]{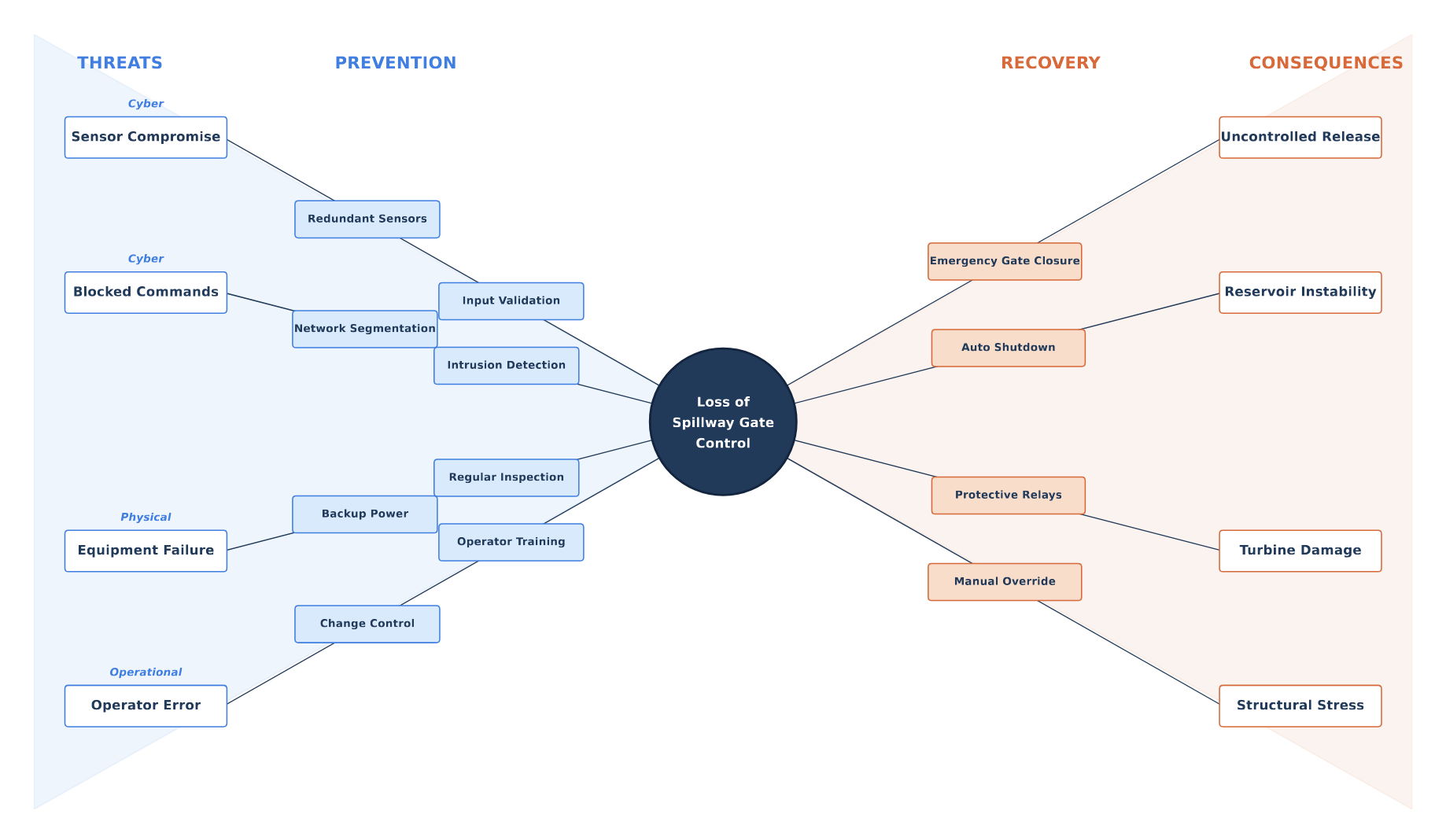}
    \centering
    \caption{BowTie diagram for loss of spillway gate control, showing threats (left), preventive barriers, the top event (centre), mitigative barriers, and consequences (right)}
    \label{images:bowtie-spillway}
\end{figure}

Fig.~\ref{images:bowtie-spillway} presents a summary BowTie diagram for loss of spillway gate control, integrating the threats, preventive barriers, consequences, and mitigative barriers discussed above. 
It groups related barriers under representative labels; the complete set of preventive and mitigative barriers is detailed in Tables~\ref{tab:bowtie-preventive} and~\ref{tab:bowtie-mitigative}.

The diagram illustrates how threats from cyber, physical, and operational domains converge on a single top event, and how layered barriers on both sides of the event provide defense in depth.

\subsection{Integration with HAZOP Findings}

The BowTie analysis demonstrates a natural connection to the HAZOP study presented in Section~\ref{sec:hazop-analysis}.
The deviations identified in the HAZOP worksheet map directly to the threats on the left side of the BowTie, as shown below. 

\begin{itemize}[nosep]
    \item The HAZOP deviation \textit{``NO level signal received by the controller''} corresponds to the BowTie threat of sensor failure or communication loss.
    \item The deviation \textit{``Control command, REVERSE''} (command direction inverted) corresponds to threats involving software faults, configuration errors, or malicious command injection.
    \item The deviation \textit{``NO status reported to the operator''} corresponds to threats that could disable monitoring and mask abnormal conditions.
\end{itemize}

Similarly, the safeguards identified in the HAZOP analysis inform the barrier identification in the BowTie.
Preventive barriers such as redundant sensors, watchdog timers, and input validation appear in both analysis, while mitigative barriers such as manual override capability and emergency procedures address the consequences identified in both methods.

This integration demonstrates how HAZOP and BowTie can be used as complementary techniques: HAZOP provides systematic identification of deviations and their causes, while BowTie organizes these findings into a visual barrier model that supports risk communication and management decision-making.

\subsection{Cyber Considerations in the BowTie Framework}

The BowTie analysis presented above includes cyber threats in addition to traditional physical and operational threats. 
Threat categories, preventive barriers, and escalation factors all account for the possibility of deliberate interference.
However, when it comes to cyber threats, the underlying barrier model on which the BowTie is built relies on assumptions that need careful consideration.

Traditional BowTie analysis assumes that the barriers are independent: the failure of one barrier does not affect the reliability of another. 
This is a reasonable assumption for physical systems where, for example, a backup power supply and a redundant sensor do not have a common failure mode. 
It is a much less reasonable assumption when the threat is an adversary with network access. 
An attacker who compromises the control network may be in a position to defeat the intrusion detection system (a preventive barrier in Table~\ref{tab:bowtie-preventive}), inject false sensor data (bypassing input validation) and suppress operator alarms (degrading a mitigation barrier in Table~\ref{tab:bowtie-mitigative}), all through the same initial foothold.
Barriers appear independent of the diagram, but share a common vulnerability: the network infrastructure on which they depend.
The escalation factors discussed above, particularly cyber-induced escalation, are not edge cases to be noted and ignored.
The pose a significant challenge to the defense-in-depth model that the BowTie structure relies upon.

A related concern is detection. 
Physical failures tend to produce observable symptoms such as: a jammed gate making noise, a blown fuse is visible, and a sensor that stops reporting triggers a signal loss alarm. 
In contrary, cyber intrusions can be designed to remain invisible.
An attacker could manipulate sensor readings gradually, staying within the plausibility bounds that would trigger conventional out-of-range alarms, while steadily moving the system toward an unsafe state.
In such a scenario, the mitigation barriers on the right side of the BowTie, many of which depend on timely operator awareness, would be less effective than the diagram suggests.

These observations do not invalidate the BowTie analysis.
But they do suggest that the barrier assessment needs to be complemented with an analysis of common-mode cyber vulnerabilities, particularly shared network dependencies and shared trust assumptions, that traditional BowTie practice does not typically examine.
For the spillway gate control subsystem, this means evaluating whether barriers that appear independent on the BowTie diagram actually rely on shared communication infrastructure, shared authentication mechanisms, or shared software components that an adversary could target as a single point of compromise.

\section{Analysis comparison between Hazop and BowTie}
\label{sec:comparitive_analysis}

Table~\ref{tab:method-comparison} highlights that HAZOP and BowTie offer complementary perspectives for analyzing the spillway gate control subsystem. 
HAZOP provides a detailed and systematic approach by analyzing each signal through guide words such as NO, MORE, and LESS, and documenting deviations together with their causes, consequences, and safeguards.
As we can see in the Table~\ref{tab:method-comparison}, the main strength of HAZOP lies in its exhaustive structure and high level of detail, which makes it possible to identify direct and indirect consequences.
However, it also indicates that HAZOP analyzes each deviation in isolation, that limits its ability to highlight coordinated causes.
It is particularly important in cyber extension cases, where an attacker may combine actions such as falsifying sensor data and suppressing alarms in order to bypass safeguards designed for independent failures.

In contrary, BowTie shows risk through a diagram that maps threats, barriers, and consequences around an unwanted event.
As we can see in Table~\ref{tab:method-comparison}, its main strength is that it is visual representation, while also making preventive and mitigative barriers explicit.
This offers a clear view of how threats progress toward a top event and how consequences unfold afterward.
At the same time, BowTie assumes barriers fail independently and may not show shared dependencies. 
In the cyber extension, this becomes significant because barriers that appear separate may rely on the same network infrastructure, allowing a single compromise to weaken multiple layers of protection. 
Overall, the comparison highlights that HAZOP is more suitable for detailed deviation analysis, while BowTie is more suitable for communicating threat pathways and barrier-based risk controls.

\begin{table*}[ht]
\centering
\caption{Comparative summary of HAZOP and BowTie applied to the spillway gate control subsystem.}
\label{tab:method-comparison}
\footnotesize
\setlength{\tabcolsep}{8pt}
\renewcommand{\arraystretch}{1.8}
\begin{tabular}{>{\raggedright\arraybackslash}p{3.5cm}
                >{\raggedright\arraybackslash}p{6cm}
                >{\raggedright\arraybackslash}p{6cm}}
\hline
\textbf{Characteristic} & \textbf{HAZOP} & \textbf{BowTie} \\
\hline
Ease of representation & Table of deviations with causes, consequences, and safeguards; Can be detailed & Diagram mapping threats, barriers, and consequences around a central unwanted event \\
Approach & Examines each signal or measurement using guide words (NO, MORE, LESS, etc.) & Maps threat-to-consequence pathways with preventive and mitigative barriers \\
Strength & Systematic and exhaustive; surfaces unlikely deviations & Visual and communicable; shows layered defense in one view \\
Limitation & Each deviation analyzed in isolation; cannot express coordinated causes & Assumes barriers fail independently; does not reveal shared dependencies \\
Cyber extension finding & Adversary can coordinate deviations (e.g., fake sensor data while suppressing alarms) to bypass safeguards designed for independent failures & Barriers that appear independent may share network infrastructure, allowing one compromise to weaken multiple layers \\
Level of detail & High and systematic & Moderate and scenario centered \\
Treatment of causes & Detailed and multiple causes can be identified & Captures threats leading to top event \\
Treatment of consequences & Captures direct and indirect consequences & Organizes consequences clearly after top event \\
Barrier representation & Listed as safeguards & Explicit preventive and mitigative barrier \\
\hline
\end{tabular}
\end{table*}

\section{Discussion}
\label{sec:discussion}

The findings of the comparisons of the methodology are based on evidence rather than opinions. 
The most striking thing about applying both methods to the same subsystem is how different the analysis looks and how they complement each other.
 
The structure of the guide word of HAZOP helps in asking questions and covering every parameter under different types of deviation, whether or not it seems intuitive.
As we can see in Table~\ref{tab:hazop-spillway}, some of the rows might not have come up in an unstructured brainstorming session, but systematic walk-through surfaces them. 
That thoroughness is the method's real value.
However, it does not enable us to identify what happens when two things go wrong at the same time, that is, each row is alone.
The cyber extension changed that: the coordinated scenario in Table~\ref{tab:cyber-hazop-demo}, where sensor manipulation and alarm suppression occur together, is not something the traditional table would ever produce, because the table has no way to express relationships between rows.
 
The BowTie works differently.
It gives you a single picture of the whole threat-barrier-consequence landscape, and that picture is genuinely useful for talking to people who are not risk specialists. 
An operator, a plant manager, or a regulator can look at Fig~\ref{images:bowtie-spillway} and understand the basic logic of how the system is protected. 
But the picture can also be misleading.
When barriers are drawn as separate boxes on the diagram, it is natural to assume that they fail independently, as a mechanical interlock and a backup power supply would. 
The cyber extension pushed back on that assumption.
If the intrusion detection system, the input validation logic, and the operator alarm display run on the same network, then a single compromise can degrade all three. 
The diagram does not show this, and nothing in the traditional BowTie process requires you to look for it.
 
The interesting part is where the two methods meet.
The HAZOP generates the raw material, the deviations, and their causes, which the BowTie then organizes into a barrier framework.
Without HAZOP, BowTie threats would be less systematically identified.
Without the BowTie, the safeguards of HAZOP would be a simple list without any visual connection to the consequences.
The cyber extension adds a third dimension to both: it asks whether the causes can be coordinated (which the HAZOP does not naturally consider) and whether the barriers are truly independent (which the BowTie does not naturally verify). 
These insights do not come from any single step in the process.
It emerges from the application of both methods and then from questioning the assumptions made by each.
 
From a practical point of view, there is something appealing about the two-stage approach. 
It does not ask practitioners to learn an entirely new methodology. HAZOP and BowTie are already familiar in process safety, and many hydropower operators will have encountered one or both.
The cyber extension layers on top of the traditional analysis rather than replacing it, means that the safety work retains its value.
For organizations where the process safety team and the cybersecurity team operate separately, this kind of shared analytical framework could be a useful starting point for closer collaboration.
As with the digitalization of these critical infrastructures, it is important to consider cyber and safety risk together and analyze how cyber risks can cause safety issues.
 
It is clear from the analysis that the HAZOP does not help in identifying the likelihood of a particular attack scenario, but it helps to identify what could go wrong and what safeguards matter.
The BowTie maps the coverage of the barrier, but does not quantify how much protection each barrier actually provides.
For decisions about where to spend limited security budgets, a quantitative layer would need to sit on top of what this paper establishes.
Currently, our analysis covers only one subsystem. 
The spillway gate control loop was chosen because it is representative, but the turbine governor or excitation system might present different risk patterns and different kinds of cyber-physical interaction that this work does not address.

\section{Conclusion and Future Work}
\label{sec:conclusion}
 
This paper applied HAZOP and BowTie analysis to the spillway gate control subsystem of a smart hydropower plant, first in traditional form and then with explicit extension to account for cyber-induced causes. The traditional analyses identified the expected range of accidental hazards and mapped the barrier framework that protects against them. 
The cyber extensions revealed that the assumptions underlying both methods, independent causes in HAZOP and independent barriers in BowTie, do not hold well against a coordinated adversary operating within the control network. 
The two methods are complementary: HAZOP systematically enumerates deviations that BowTie then organizes into a visual barrier model, and the cyber extension applied to both exposes risks that neither method reveals in its traditional form.
 
Several directions for future work follow. 
A full cyber-extended HAZOP covering all 18 deviations would provide a more complete threat picture and likely reveal additional coordination patterns beyond the two-row demonstration shown here.
Applying the same two-stage approach to other subsystems, such as the turbine governor or the excitation system, would test whether the findings are generalizable across different control architectures. 
Integrating a structured threat modeling framework such as STRIDE for ICS could provide a more systematic basis for identifying cyber-induced causes, moving beyond the analyst-driven approach used here. Quantitative risk assessment, based on the qualitative foundation this document establishes, would support the prioritization of security investments in operational settings. 
Finally, applying additional methodologies such as CORAS to the same subsystem would enable a comparative evaluation of how different risk assessment approaches perform in the hydropower domain, extending the multi-method comparison.
Moreover, all these risk analysis methods do not consider exploitability, that is, how easy it is to target specific system.
Furthermore, it is also important to consider the level of attackers, i.e., how skilled they are, which has not been addressed in these methodologies.






\bibliographystyle{elsarticle-num}
\small
\bibliography{references}



\end{document}